\documentclass[aps,pra,twocolumn,nofootinbib,amsmath,amssymb,superscriptaddress]{revtex4-1}
\usepackage{graphicx}% Include figure files
\usepackage{dcolumn}% Align table columns on decimal point
\usepackage{bm}% bold math

\begin{document}

\title{Anti-mirror reflection of a bounded plane optical waveguide: string model}

\author{P. Yu. Shapochkin}
\author{Yu. V. Kapitonov}
\email[]{kapiton22@gmail.com}
\affiliation{St. Petersburg State University, Physics department, 198504 St. Petersburg, Russia}
\author{G. G. Kozlov}
\affiliation{St. Petersburg State University, Physics department, 198504 St. Petersburg, Russia}
\affiliation{St. Petersburg State University, Spin Optics Laboratory, 198504 St. Petersburg, Russia}

\date{\today}

\begin{abstract}
The effect of anti-mirror reflection from a bounded plane optical waveguide used earlier for observation of slow light in a Bragg waveguide is considered. A theory of the effect is proposed that allowed us to interpret the observed spectral (Gaussian lineshape) and angular (Lorentzian angular profile) properties of the light scattered  by the bounded waveguide in the anti-specular direction. 
\end{abstract}

% insert suggested PACS numbers in braces on next line
% insert suggested keywords - APS authors don't need to do this

\maketitle

\section{Introduction}

The plane optical waveguide (POW) comprised of two parallel plane reflectors separated by a waveguiding gap (Fabry-Perot interferometer) plays important role in optics and is studied in great details \cite{B&W}.  Still, the interest to this simple resonant system has considerably risen during the last years. This is mainly connected with the development of technology of small-sized POW (microcavities) with the possibility to place structures with material resonances (quantum wells and quantum dots) into the microcavity gap, whose interaction with optical mode of the microcavity leads to a number of interesting phenomena \cite{Kavokin}. At the same time, the problems arising in the up-to-date technologies make us to recall about known properties of the empty  POW with the aim to use them in the optical systems of information processing. In particular, in the recent papers \cite{Japan,Kozlov}, it was proposed to use POW for obtaining the so-called "slow light, and the possibility of application of such a waveguide in optical switching and buffering systems has been noted. In \cite{Kozlov}, the light slow down in POW has been demonstrated experimentally with the use of the phenomenon of anti-mirror reflection arising upon light scattering in a {\it bounded} POW. In this paper, we propose a model suitable for quantitative description of the anti-mirror reflection, and, in the framework of this model, we interpret the angular and spectral characteristics of this reflection.

\section{Anti-mirror reflection: the string model}

Recall, in brief, the mechanism of anti-mirror reflection upon excitation of a bounded POW by a plane monochromatic wave \cite{Kozlov}. First, consider, for this purpose, an {\it infinite} POW. In this case, the incident wave, generally, is efficiently reflected by the mirror upon which it falls (referred hereafter to as {\it top} with the other called {\it bottom}). As a result, the field strength, in the waveguiding gap, appears to be small and the POW reflectivity close to unity.  This picture of reflection is broken at resonance, when the phase increment of the wave, for the round-trip over the POW, is a integer multiple of $2\pi$. Under this condition, the field strength in the waveguiding gap resonantly increases, and, in spite of the fact that the waveguiding gap is separated from the surrounding space by highly reflecting mirrors, the field inside this gap starts to play a decisive role in formation of the reflected and transmitted waves. Under resonance conditions, the field inside the waveguiding gap may be considered as a quasi-one-dimensional wave of large amplitude with the wave number equal to the projection $k_x$ of the incident wave vector upon the POW plane. In virtue of imperfection of the POW mirrors, propagation of this wave occurs with leakage, which, due to the high field strength in the waveguiding gap, provides the resonant transmission coefficient of the POW close to unity.  As for the reflected wave, its amplitude is determined by interference of the wave arisen due to the leakage from the guiding gap and the wave reflected by the top mirror. In conformity with the energy conservation law, this interference is destructive, thus leading to resonant reduction of the POW reflectivity.

\begin{figure}
\includegraphics[width=0.8\columnwidth,clip]{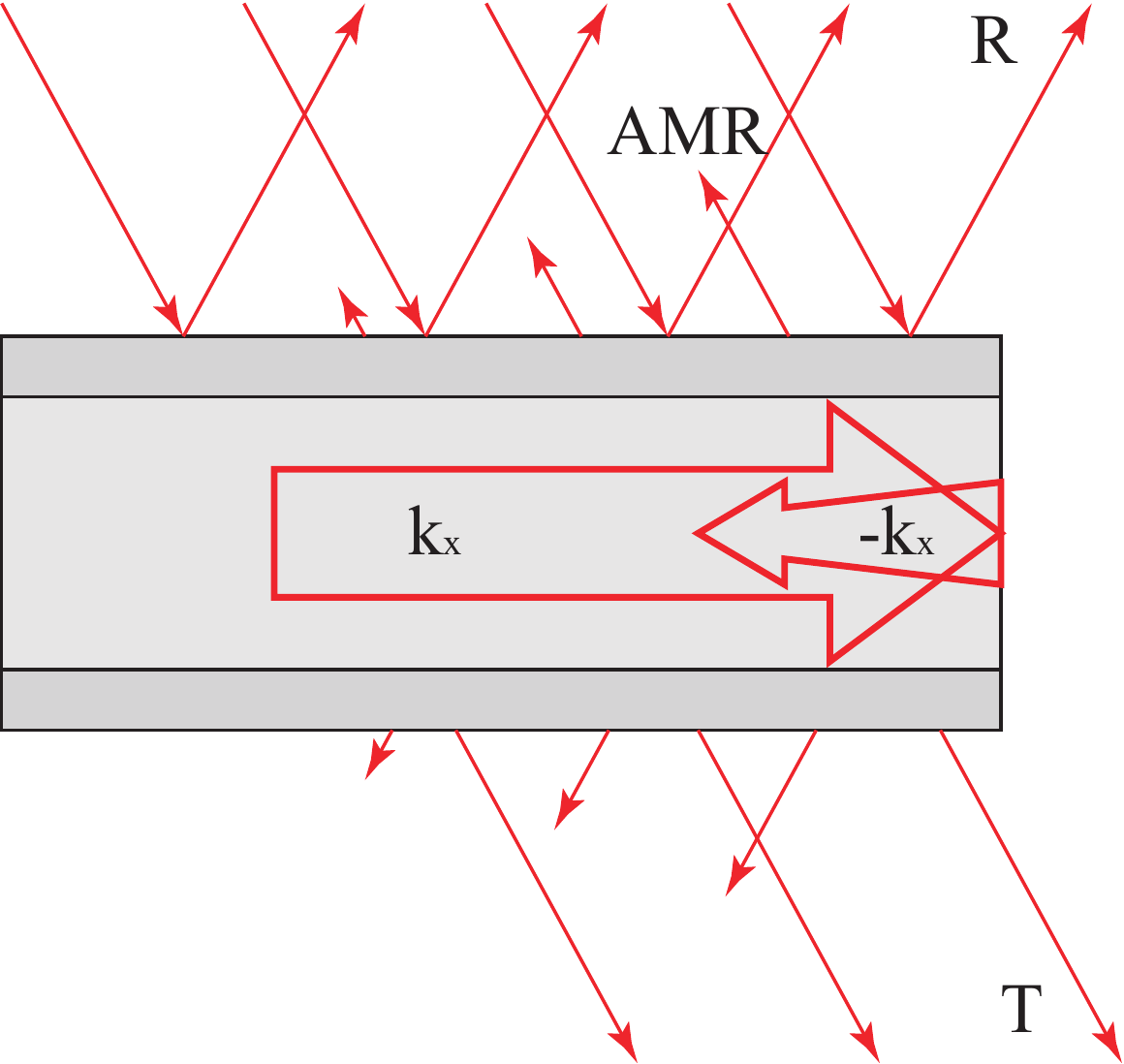}
\caption{\label{Fig1}Formation of the anti-mirror reflection in the semi-infinite waveguide. R - reflection, T - transmission, AMR - anti-mirror reflection.}
\end{figure}

The above consideration is valid for {\it infinite} POWs. Assume that the POW under study is {\it semi-infinite}, with the resonant plane monochromatic wave incident upon it. In the waveguiding gap of the semi-infinite POW, like in the case of infinite POW, there will arise the wave of large amplitude with the wave number $k_x$ (big right arrow in Fig.~\ref{Fig1}). Specific feature of the semi-infinite waveguide is that this wave should be inevitably reflected by the waveguide bound giving rise to the wave with the wave number $-k_x$ (small left arrow in Fig. 1). The leakage of this wave gives rise to the anti-mirror wave propagating in the backward direction \cite{Kozlov}. The effect of anti-mirror reflection should accompany the reflection from an arbitrary bounded plane structure with a surface electromagnetic mode. Layered structure studied in \cite{Soboleva} can serve as  an example. Note, finally,  that a similar effect of the anti-mirror reflection in plane crystal systems with excitonic type of susceptibility has been described in \cite{Kozlov1}.

Since in the capacity of reflectors, in the present-day optical POW, are used multi-layer Bragg mirrors, a rigorous analysis of the anti-mirror reflection requires solution of the scattering problem on a semi-infinite layered structure and can hardly be performed analytically.  This is why, to quantitatively describe this phenomenon, we propose here a simple {\it string model} based on the qualitative picture of scattering on the POW-type structures described above. In the proposed model, the plane optical waveguide is modeled by an effective string excited by an external field whose spatial distribution corresponds to the field incident upon the optical waveguide. Basic statements of the model are:

\begin{enumerate}
\item
The waveguide is replaced by an effective string, with the wave dispersion law $Q(\omega)$ chosen the same as for the waves in the POW. This dispersion law is implicitly defined by the relationship

\begin{equation}
 \bigg ({\omega n_2\over c}\bigg )^2-Q^2=\bigg ({\pi m\over 2L}-{\imath \hbox{ ln }r\over 4L}\bigg )^2.
\label{51}
\end{equation}

Here $Q$ is the wave number, $\omega$ is the optical field frequency, $c$ is speed of light, $2L$ is the waveguide gap width, $n_2$ is the refractive index of material in the waveguiding gap, $r\equiv r_{2\rightarrow 1} r_{2\rightarrow 3}$ is the product of the reflectivities of the top and bottom mirrors for the light incident from the side of the guiding gap (module of this parameter for the up-to-date reflecting mirrors is close to unity), and $m$ is the POW mode number. 

\item
The POW plane is assumed to be the $xy$-plane, and the fields under consideration to be independent on $y$.  Then the wave operator entering the equation for displacement of the effective string can be obtained from ({\ref{51}}) by the substitution $Q\rightarrow -\imath \partial/\partial x$. All the time dependences are implied to be proportional to $e^{-\imath\omega t}$. 

\item
We assume that the effective string corresponding to the POW is excited by a distributed external force, whose spatial dependence is controlled by the incident field $A(x)$ and write the equation for the effective string displacement in the form:

\begin{eqnarray}
{\partial^2 U\over\partial x^2}+K^2 U= A(x),
\nonumber\\
K^2\equiv\bigg ({n_2\omega\over c}\bigg )^2 -
\bigg ({\pi m\over 2L}-{\imath \hbox{ ln }r\over 4L}\bigg )^2.
\label{52}
\end{eqnarray}

\item
The reflected $R(x)$ and transmitted $T(x)$ fields in the POW plane can be calculated through displacement of the effective string as

\begin{eqnarray}
R(x)=\alpha A(x)+\beta U(x),
\nonumber\\
T(x)=\gamma U(x).
\label{53}
\end{eqnarray}

With the coefficients $\alpha$, $\beta$, and $\gamma$ chosen so that the reflection and transmission spectra obtained in the framework of the string model using (\ref{52}) and (\ref{53}) coincide with those for the exactly solvable case of infinite waveguide with the plane wave incident at the angle $\phi$.  Such a coincidence (at frequencies close to the POW resonant frequency) appears to be possible for the following values of the coefficients $\alpha$, $\beta$, and $\gamma$:

\begin{eqnarray}
\gamma={\imath \pi m\over 4L^2} \hskip2mm t_{1\rightarrow 2}t_{2\rightarrow 3} \hskip2mm e^{-\imath [ k_z+p_z -2g_z]L},
\nonumber\\
\alpha=r_{1\rightarrow 2}e^{-2\imath k_zL} ,
\nonumber\\
\beta={\imath \pi m\over 4L^2} \hskip2mm  t_{1\rightarrow 2}t_{2\rightarrow 1}r_{2\rightarrow 3} \hskip2mm e^{-2\imath k_zL},
\nonumber\\
k\equiv{\omega n_1\over c},
\nonumber\\
k_z\equiv k\cos\phi,
\nonumber\\
g_z^2\equiv \bigg ( {\omega\over c}\bigg )^2\bigg (n_2^2-n_1^2\sin^2\phi \bigg ),
\nonumber\\
p_z^2\equiv \bigg ( {\omega\over c}\bigg )^2\bigg (n_3^2-n_1^2\sin^2\phi \bigg ).
\label{57}
\end{eqnarray}

Where $n_1$ and $n_3$ are the refractive indices of the media above and below the POW, respectively, $t_{1\rightarrow 2}$ and $t_{2\rightarrow 1}$ are the transmission coefficients of the top mirror for the light incident from the upper half-space and from the waveguiding gap, respectively, $t_{2\rightarrow 3}$ is the transmission coefficient of the bottom mirror for the light incident from the side of waveguiding gap, $r_{1\rightarrow 2}$ is the reflectivity of the top mirror for the light incident from the upper half-space.  All these coefficients can be expressed in terms of the elements of transfer matrices of the used (e.g., Bragg) mirrors. If the incident field is not characterized by a certain angle of incidence, but represents a sum of plane-wave constituents with the angles of incidence lying within the range $[\phi-\Delta\phi, \phi+\Delta\phi]$, then the above expressions, for $\Delta\phi/\phi<1$ can still be used for evaluation. 
 
\item
The reflected and transmitted fields can be found by solving the wave equation in the form

\begin{equation}
\bigg ({\partial^2\over \partial x^2}+{\partial^2\over \partial z^2}\bigg )v(x,z)=
- \bigg ({\omega n_i\over c}\bigg )^2 v(x,z)
\label{54}
\end{equation}

at $i = 1,3$.  To find the reflected (transmitted) field in the upper (lower) half-space, this equation should be solved for $z > 0$ ($z < 0$) with the boundary condition $v(x,0)=R(x)$  ($v(x,0)=T(x)$), assuming $i=1$ ($i=3$).   The functions $R(x)$ and $T(x)$ can be found from (\ref{52}) and (\ref{53}).  Note that the above model can be extended to the case when wave propagation in the POW is two-dimensional. This corresponds to passing from (\ref{52}) to the following two-dimensional equation:

\begin{equation}
{\partial^2 U\over\partial x^2}+
{\partial^2 U\over\partial y^2}
+\bigg ({n_2\omega\over c}\bigg )^2 U
-\bigg ({\pi m\over 2L}-{\imath \hbox{ ln }r\over 4L}\bigg )^2U= A(x).
\label{2d}
\end{equation}

\end{enumerate}

\subsection{Anti-mirror redlection for the case of plane wave}

To analyze the anti-mirror reflection using the string model, we assume this string to be semi-infinite and occupying positive semi-axis $x\in [0,\infty]$. As a boundary condition at $x=0$, we assume  $U(x=0)=0$, i.e., the field strength in the waveguiding gap vanishes at the end of the waveguide. Our failures to inject the light into the waveguide through its butt-end \cite{Kozlov} confirm this assumption. Thus, we have to find solution of (\ref{52}) under conditions that $U(0)=0$ and function $U(x)$ is bounded at $x>0$. In this section we will consider the case when the exciting field is a plane wave $A(x)\rightarrow Ae^{\imath qx}$, with $q=\omega n_1\sin\phi/c$, $\phi$ is the angle of incidence. General solution of inhomogeneous equation (\ref{52}) may be represented as a sum of its particular solution and the general solution of the corresponding homogeneous equation:

\begin{eqnarray}
{\partial^2 U\over\partial x^2}=-K^2U,
\hskip5mm
K= K'+\imath K'',
\nonumber\\
K'\approx\sqrt{\bigg ({n_2\omega\over c}\bigg )^2 - \bigg ({\pi m\over 2L}\bigg )^2},
\nonumber\\
K''\approx {\pi m \hbox { ln } r\over 4 K' L^2}<0.
\label{58}
\end{eqnarray}

Approximate equalities correspond to smallness of ln $r$ which will be assumed hereafter \footnote{This assumption corresponds to the case of POW with highly reflecting mirrors. In the up-to-date POW, $|$ln $r| < 10^{-3}$.}. Then, the general solution of (\ref{58}), bounded at $x>0$, will have the form:

\begin{equation}
U_h(x)=U_{h0}e^{-\imath Kx}.
\label{59}
\end{equation}

The solution $\sim e^{\imath Kx}$ is omitted, since it increases at $x\rightarrow +\infty$ due to negative imaginary part of $K$. Particular solution of inhomogeneous equation (\ref{52}) with its right-hand side in the form  $Ae^{\imath q x}$ can be written as follows:

\begin{equation}
U_i(x)=\Theta(x) A \hskip2mm {e^{ \imath qx}\over K^2-q^2}.
\label{60}
\end{equation}

By selecting $U_{h0}$ so that the sum of the homogeneous and inhomogeneous solutions vanishes at $x=0$, we obtain that solution of (\ref{58}) meeting the above conditions has the form:

\begin{equation}
U(x)=\Theta(x)A \hskip2mm {e^{\imath qx}-e^{-\imath Kx}\over K^2-q^2}.
\label{61}
\end{equation}

The first term in the nominator describes the usual mirror reflection with the same $x$-projection of the wave vector as that of the incident wave. The anti-mirror reflection is described by the last term in the nominator, and below we will consider only its contribution into the field in the upper half-space. To find this contribution, we have to solve (\ref{54}) with the appropriate boundary condition:

\begin{eqnarray}
\bigg ({\partial^2\over \partial x^2}+{\partial^2\over \partial z^2}\bigg )v(x,z)=
- \bigg ({\omega n_1\over c}\bigg )^2 v(x,z),
\nonumber\\
v(x,0)=-\beta A \hskip1mm {\Theta(x)\over K^2-q^2} \hskip2mm e^{-\imath Kx}.
\label{62}
\end{eqnarray}

The solution is sought in the form of a sum of plane waves in the upper half-space:

\begin{equation}
v(x,z)=\int  V_pe^{\imath [px-\sqrt{k^2-p^2}z]}\hskip1mm dp.
\end{equation}

The boundary condition (\ref{62}) allows us to obtain explicitly the field $v(x,z)$ and its spatial harmonics $V_p$ using the following chain of relationships:

\begin{eqnarray}
-\beta A\hskip1mm {\Theta(x)\over K^2-q^2} \hskip2mm e^{-\imath Kx}=
\int  V_pe^{\imath px}\hskip1mm dp \hskip5mm\Rightarrow
\nonumber\\
V_p=-{1\over 2\pi}\hskip2mm
{\beta A\over K^2-q^2} \int_0^\infty  e^{-\imath (K+p)x} dx
\hskip5mm\Rightarrow
\nonumber\\
V_p={\imath\over 2\pi}\hskip2mm
{\beta A\over K^2-q^2} \hskip2mm {1\over K+p }\hskip2mm\Rightarrow
\nonumber\\
v(x,z)=
{\imath\over 2\pi}\hskip2mm
{\beta A\over K^2-q^2}
\int  {e^{\imath [px-\sqrt{k^2-p^2}z]}\over K+p}\hskip1mm dp.
\end{eqnarray}

It is seen from the above expression that in the field scattered along the anti-specular direction, there prevail plane-wave components with $p\sim -K'$ (the point $p=-K'$ is the closest to the pole of the integrant). Amplitude of the scattered field is the greatest at $|K'|=|q|$, which corresponds to the condition of resonant excitation of the POW.

\begin{figure}
\includegraphics[width=1\columnwidth,clip]{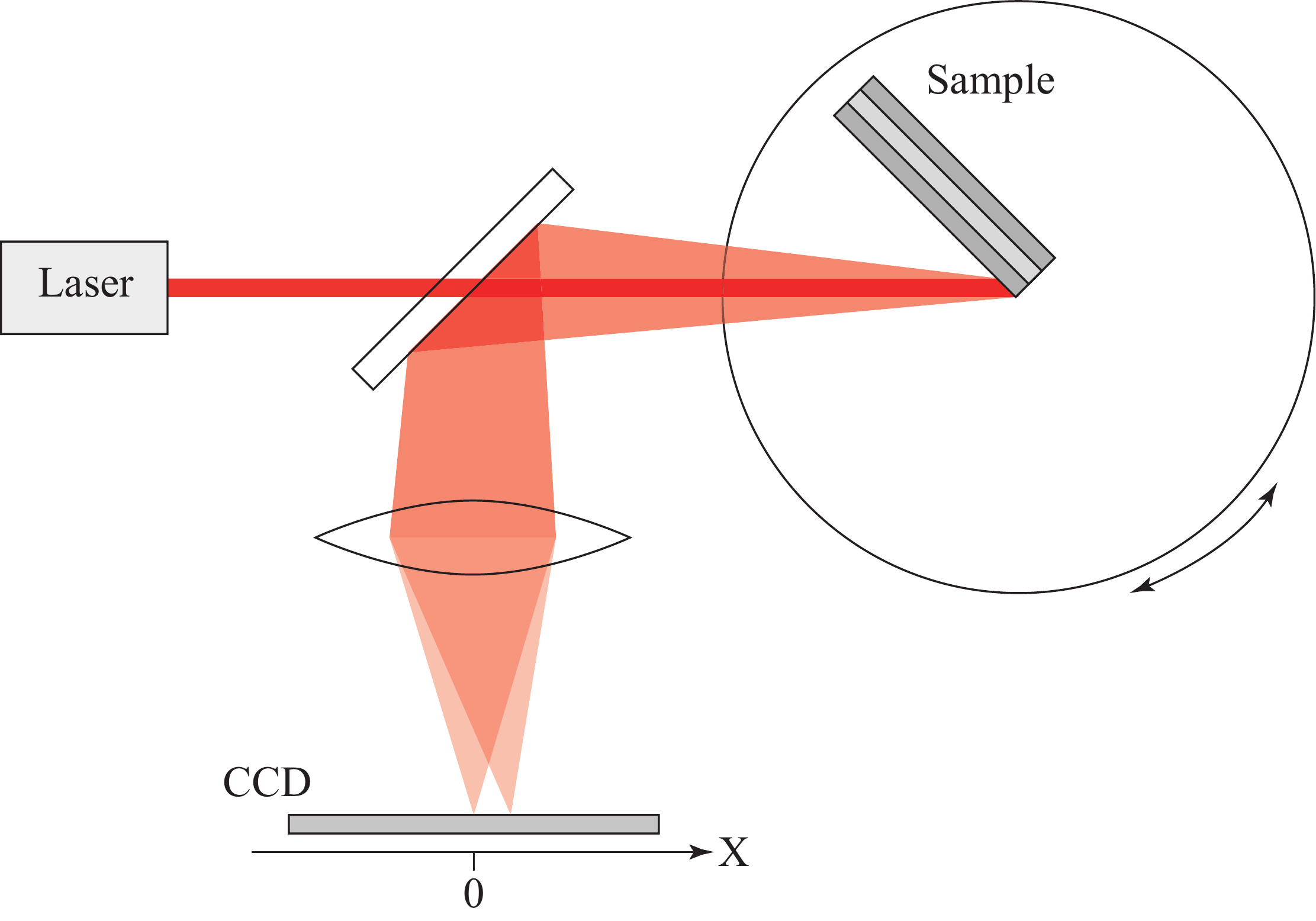}
\caption{\label{Fig2}Schematic of the experiment aimed at observation of angular distribution of the light scattered in the anti-specular direction}
\end{figure}

In the experimental detection of angular distribution of the anti-reflected field (Fig.~\ref{Fig2}), the light scattered in the anti-specular direction is collected by the lens, with a CCD matrix placed in its focal plane. 

If the lens is aligned so that the plane wave with the $x$-projection of the wave vector equals to $-q$ (anti-specular with respect to the incident one) is collected by the lens into central pixel of the CCD matrix, whose coordinate is taken to be zero (Fig.~\ref{Fig2}), then the intensity distribution in the plane of the CCD matrix will  correspond to squares of modules of the harmonics $|V_p|^2$, with the intensity at the central pixel  being $|V_{-q}|^2$. Assuming that angular width of the anti-reflected beam is small and expressing the coordinate of the pixel, into which the harmonics with the $x$-projection of the wave vector $p$ is focused, in termes of the focal length $f$ of the used lens, we find that the distribution of brightness $S(X)$ detected in the CCD plane has the form:

\begin{eqnarray}
S(X)& = & |V_p|^2\bigg |_{p=-q+k \cos\phi X/f} = \bigg |{\beta A\over 2\pi}\bigg |^2
 	\bigg |{1\over K^2-q^2}\bigg |^2 \times
 	\nonumber \\
 	& \times & 	{1\over (K'-q+k\cos\phi \hskip1mm X/f)^2+(K'')^2}.
 	\label{67}
\end{eqnarray}

where $q=[\omega n_1/c]\sin\phi$.  Thus, observing the function $S(X)$ by means of the above method at different $\omega$, we should obtain the function with Lorentzian profile, with its amplitude attaining maximum when $\omega$  appears to be resonant for the given angle of incidence \footnote{This frequency can be found from the condition $K'=q$:
$$\omega_{res}={\pi c m\over 2L\sqrt{n_2^2-n_1^2\sin^2\phi}}.$$}.

\begin{figure}
\includegraphics[width=1\columnwidth,clip]{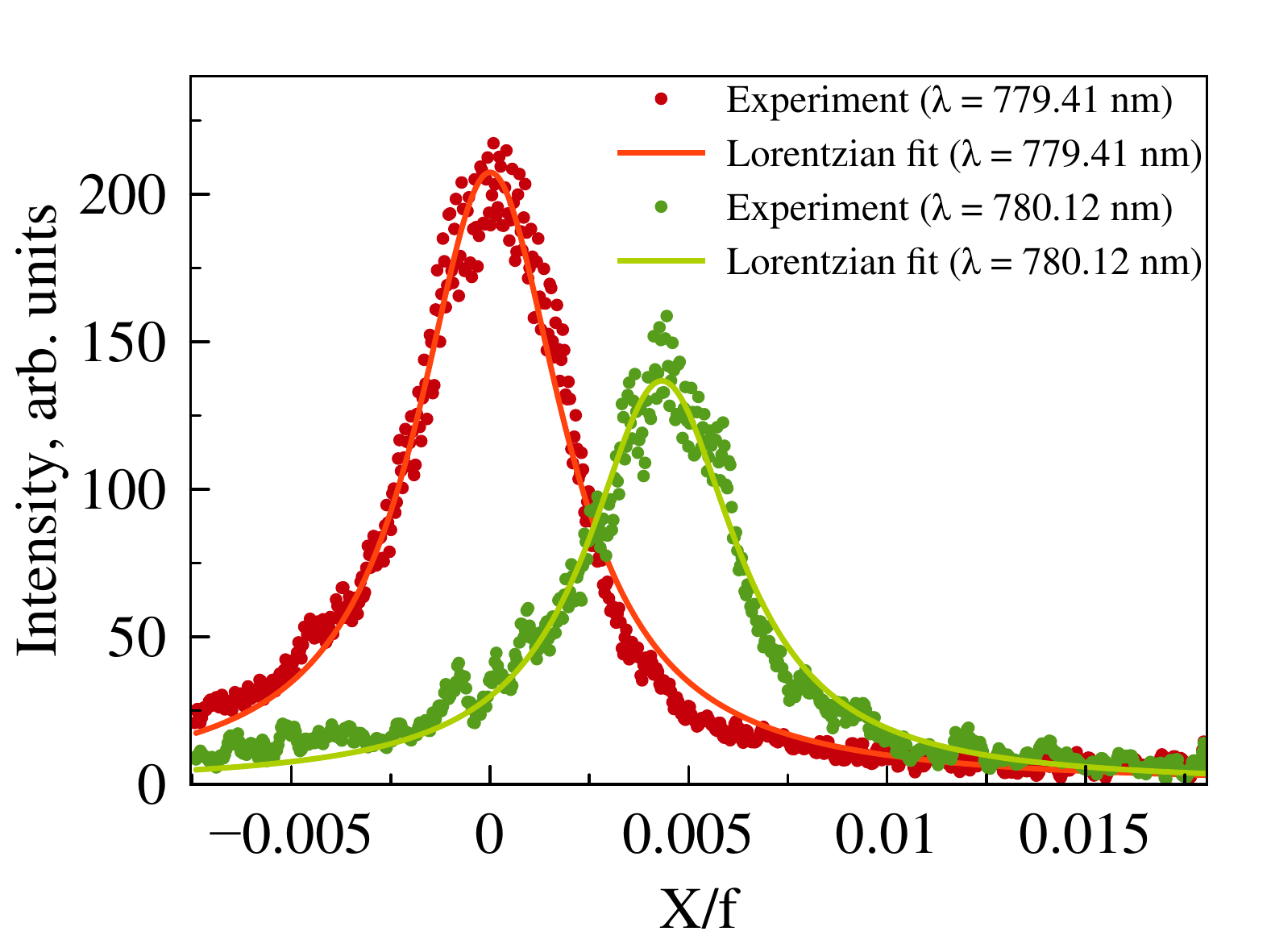}
\caption{\label{Fig3}Angular distribution of the light scattered by the POW in the anti-specular direction at exact resonance (left curve) and under slight detuning (right curve). At resonance, angular distribution of the scattered light is peaked exactly in the anti-specular direction. When moving away from the resonance, amplitude of the angular distribution decreases, while its peak shifts in accordance with (\ref{67})}
\end{figure}

In this case the peak of this curve passes through the center of the CCD matrix  shifting away from the center upon frequency detuning from the resonance. This is exactly what we observed in the experiment (Fig.~\ref{Fig3}). 

\subsection{Anti-mirror reflection for the case of nonuniform illumination}

\begin{figure}
\includegraphics[width=1\columnwidth,clip]{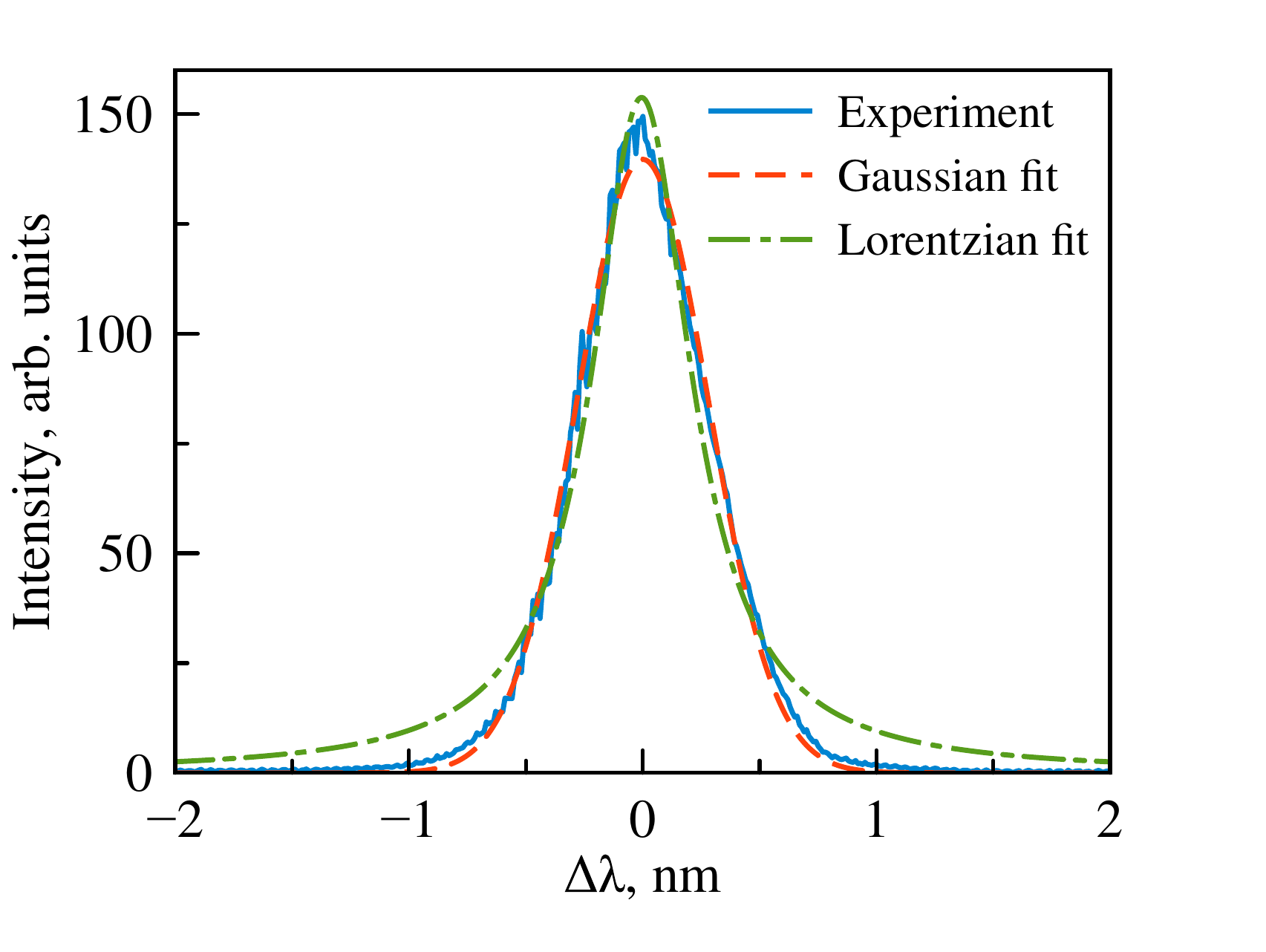}
\caption{\label{Fig4}The anti-mirror reflection spectrum and its approximation by the Lorentzian and Gaussian peaks}
\end{figure}

To increase intensity of the anti-mirror reflection from a bounded POW, is convenient to focus the incident beam. Measurements of the anti-mirror reflection spectra in these experiments often lead to {\it Gaussian} shape of the resonant curve (Fig.~\ref{Fig4}), which is unusual for spectroscopy of optical cavities. For instance, the conventional reflection spectrum of POW upon its excitation by a plane wave has {\it Lorentzian} shape \footnote{Deformation of the spectrum upon focusing of the incident beam is also possible for the unbounded POW.}. Below, we present interpretation of this reshaping of the spectrum in the framework of the string model. 

Consider the case, when the waveguide is excited not by a plane wave, but  rather by the field $a(x)$ with the frequency $\omega$ distributed, in some way, over its surface (e.g., it may be the light beam focused at some distance from the edge of the waveguide). Let us put the question about spectrum of the anti-mirror reflection or, in other words, about frequency dependence of intensity of the light scattered in the anti-specular direction.

To analyze such a problem in the framework of the string model, we have to solve (\ref{52}) when  $A(x)\rightarrow a(x)$  \footnote{Below, we consider the case when the field on the POW is nonuniform only in the direction of the $x$-axis. It corresponds to focusing by a cylindrical lens. The case of focusing in two directions will be briefly described in the next section.}.  Let us represent the function $a(x)$ in the form of Fourier integral over plane waves:

\begin{eqnarray}
a(x)=\int A_pe^{\imath px} dp,
\nonumber\\
A_p={1\over 2\pi}\int_0^\infty a(x)e^{-\imath p x}dx.
\label{68}
\end{eqnarray}

Here, we have taken into account that the function $a(x)$ can be considered nonzero only at $x>0$, i.e., in the region where our semi-infinite waveguide is located. The function $U$ for the case of excitation by a plane wave has been obtained above (\ref{61}), and, to find function $U$ corresponding to the sum of the plane waves (\ref{68}), we have to set $A\rightarrow A_p$ in (\ref{61}) and to integrate it over $p$:

\begin{equation}
U(x)=\Theta (x)\int A_p {e^{\imath px}-e^{-\imath Kx}\over K^2-p^2} dp.
\label{69}
\end{equation}

The anti-specular part of the reflected field $R_{a}(x)$, we are interested in \footnote{In what follows, we assume that the incidence of the light beam is well far from normal, and the anti-specular beam can be easily separated from the specular one.}, is controlled by the last term in the nominator $\sim e^{-\imath Kx}$:

\begin{eqnarray}
R_a(x) =  -\Theta (x)\beta \hskip2mm e^{-\imath Kx}\hskip2mm
 \int {A_p \over K^2-p^2 }  dp =
        \nonumber \\
 =  -\Theta (x){\beta \hskip2mm e^{-\imath Kx}\over 2\pi}
 \int _0^\infty dx'\hskip1mm a(x')\int dp\hskip1mm {e^{-\imath px'}\over K^2-p^2}.
 \label{70}
\end{eqnarray}

Let us calculate the inner integral $I$ over $p$ for positive $x'$. In this case, an important role is played by the pole of the integrant in the lower half-space of the complex $p$, and the result has the form $ I= \pi\imath\hskip1mm e^{-\imath K x'}/K$, while the anti-mirror reflection spectrum $s(\omega)$, we are interested in, is given by the expression

\begin{eqnarray}
s(\omega)=H(\omega,x)\hskip1mm G(\omega),
\nonumber\\
G(\omega)\equiv\bigg |\int_0^\infty dx'\hskip1mm a(x')e^{-\imath Kx'}\bigg |^2,
\nonumber\\
H(\omega,x)\equiv \bigg |{\beta\over 2K}\bigg |^2\hskip2mm e^{2 K''x}.
\label{73}
\end{eqnarray}

One can see from this equation that this spectrum depends on $x$ \footnote{The function $s(\omega)=|R_a(x)|^2$, thus obtained, yields the intensity distribution of the fraction of the field that forms the anti-mirror reflection, and this frequency dependence may be different for different points of the POW surface}. This, however, does not bring additional difficulties because the $x$-dependent factor $H(\omega,x)$ is a weak function of frequency in the resonance region. For this reason, the frequency behavior of the anti-mirror reflection is mainly determined by the $x$-independent factor $G(\omega)$.

With the aid of the equations, thus obtained, we can interpret the experimentally observed Gaussian shape of the anti-mirror reflection spectrum.  To do this, let us consider the case when the incident beam has a Gaussian profile with the width $d$ centered at $x_0$ and falls at the angle $phi$.  In this case, the function $a(x)$ has the form 

\begin{eqnarray}
a(x)=a_0\hskip1mm \Theta (x)\hskip1mm \exp\bigg [\imath q x-\bigg ({x-x_0\over d}\bigg )^2\bigg ],
\nonumber\\
q={\omega n_1\over c}\hskip1mm \sin\phi.
\label{75a}
\end{eqnarray}

For the function $G(\omega)$ (\ref{73}), we obtain

\begin{equation}
G(\omega)=a_0^2\left |
\int_0^\infty dx
\exp\left [\imath (q-K)x-\left ({x-x_0\over d}\right )^2\right ]
\right |^2.
\label{75}
\end{equation}

If the maximum of the integrant is displaced from the point $x = 0$ at the distance exceeding the width of this peak (the corresponding condition is $ x_0+K''d^2/2>d$), then the integration can be extended to infinite limits. Now the integral can be taken analytically, and for the function $G(\omega)$ we have: 

\begin{equation}
G(\omega)=\pi a_0^2d^2
\bigg |
\exp\bigg [-
{ [K-q]^2d^2\over 4} \bigg ]
\bigg |^2\hskip2mm\exp\bigg [ 2K''x_0\bigg ]
\end{equation}

at

\begin{equation}
 x_0+K''d^2/2>d.
\label{78}
\end{equation}

At  resonance, $K'-q$ passes through zero changing its sign, and, as a result, the spectrum acquires Gaussian shape (at $|K''|<<|K'|$). At $d>> 1/|K''|$, the spectrum becomes Lorentzian. In this case, the width of the Gaussian function (\ref{75}) starts to exceed the decay length  $1/|K''|$, and, when calculating the integral (\ref{75}) we may assume that $d\rightarrow\infty$, and thus we have:

\begin{equation}
G(\omega)={a_0^2\over (q-K')^2+(K'')^2}.
\label{79}
\end{equation}

Direct numerical calculations of integral (\ref{75}) for arbitrary $x_0$ and $d$ confirm the above limiting cases.

\section {Conclusions}

In summary, we have considered the effect of anti-mirror reflection and propose the string model suitable for description of this effect. In the framework of the proposed model we analyze the angular and frequency behavior of the light scattered in the anti-specular direction. We show that the angular distribution of the anti-mirror reflection (reduced to distribution of brightness in the plane of the detecting CCD matrix) is described by Lorentzian. The anti-mirror reflection spectrum appears to be strongly dependent on properties of the incident beam, being Lorentzian for the plane-wave irradiation and Gaussian for irradiation by focused beam.  The results of calculations are illustrated by experimental data. 

It is noteworthy that the above consideration of the anti-mirror reflection under conditions of nonuniform excitation corresponds to the case of one-dimensional focusing with the aid of a {\it cylindrical} lens. Analysis of the two-dimensional focusing using {\it spherical} lens is also possible in the framework of the generalized string model and can be performed using (\ref{2d}). The result of calculation of the function $R_a$ for the case of the irradiating Gaussian beam with the widths $d_x$ and $d_y$ along the $x$ and $y$ axes in the POW plane, respectively (the sense of all the rest parameters is the same as in the previous section) has the form:

\begin{eqnarray}
R_a(x,y)& = &-{\imath a_0 d_y\over 4\sqrt\pi}\int_0^\infty dx'\int_{-\infty}^\infty dp
\exp\bigg [\imath q x'-\bigg ({x'-x_0\over d_x}\bigg )^2\bigg ] \times
        \nonumber \\
 & \times &   \exp -\bigg ({pd_y\over 2}\bigg )^2\times
 \nonumber \\
 & \times & {\exp \imath \bigg(-\sqrt{K^2-p^2}\hskip1mm (x+x')+py \bigg)\over \sqrt{K^2-p^2}}.
 \label{22d}
\end{eqnarray}

If the spectrum, upon its detection, is averaged over the POW boundary (i.e., the detected quantity is $\sim\int dy R_a(x,y)$), then the difference between the cases of spherical and cylindrical focusing disappears, since  integration of (\ref{22d}) over $y$ and calculation of the module squared leads to (\ref{75}). This is confirmed by our experiments which show that the character of the anti-mirror reflection spectrum depends weakly on which kind of lens (spherical or cylindrical) is used for the beam focusing.

\begin{acknowledgments}
The financial support from the Russian Ministry of Education and Science (Contract No. 11.G34.31.0067 with SPbSU and leading scientist A. V. Kavokin) is acknowledged. This work was carried out on the equipment of SPbU Resource Center "Nanophotonics" (photon.spbu.ru).
\end{acknowledgments}

\bibliography{AMR_revtex}

\end{document}